\documentclass{elsart1p}
\usepackage{bm, amsmath, amssymb}
\usepackage{graphicx}
\usepackage{subfigure}
\usepackage{color}
\usepackage{setspace}
\newcommand \etc {{\sl etc.\ }}
\newcommand \eg {{\sl e.g.\ }}
\newcommand \ie {{\sl i.e.\ }}
\newcommand \viz {{\sl viz.\ }}
\newcommand \ibid {{\sl ibid.\ }}
\newcommand \fig[1] {Fig.\ \ref{#1}}

\newcommand \Ref[1] {Ref.\ \cite{#1}}

\newcommand \beq {\begin{equation}}
\newcommand \eeq {\end{equation}}
\newcommand \beqa {\begin{eqnarray}}
\newcommand \eeqa {\end{eqnarray}}
\newcommand \bsubeq {\begin{subequations}}
\newcommand \esubeq {\end{subequations}}
\newcommand \benum {\begin{enumerate}}
\newcommand \eenum {\end{enumerate}}
\newcommand \bfig {\begin{figure}[!t]\begin{center}}
\newcommand \efig {\end{center}\end{figure}}
\newcommand \btab {\begin{table}[!ht]\begin{center}}
\newcommand \etab {\end{center}\end{table}}

\newcommand \prl[3] {Phys.\ Rev.\ Lett.\ {\bf #1},\ (#3)\ #2}
\newcommand \prd[3] {Phys.\ Rev.\ {\bf D#1},\ (#3)\ #2}

\newcommand \plb[3] {Phys.\ Lett.\ {\bf B#1},\ (#3)\ #2}

\newcommand \jhep[3] {JHEP\ {\bf #1},\ (#3)\ #2}

\begin{document}
\begin{frontmatter}

\title{Screening of light mesons and charmonia at high temperature}
\author{Swagato Mukherjee\thanksref{coll}} 
\address{Fakult\"at f\"ur Physik, Universit\"at Bielefeld, D-33615 Bielefeld, 
Germany.} 
\ead{smukher@physik.uni-bielefeld.de}
\thanks[coll]{ On behalf of the \emph{RBC-Bielefeld collaboration}.  The numerical
calculations presented here have been carried out on the apeNEXT computers of
Bielefeld University, the QCDOC computers at BNL and the BlueGene/L computer at the
New York Center for Computational Sciences which is supported by the U.S. Department
of Energy under Contract No.  DE-AC02-98CH10886 and by the State of New York.  }

\begin{abstract}

We present lattice QCD results for the screening masses of light mesons and
charmonia. The lattice computations were performed with $2+1$ flavors of improved
staggered quarks using quark masses which correspond to realistic pion and kaon
masses at zero temperature. For the light quark sector we have found that the
screening masses in the pseudo-scalar and the isovector scalar channels do not become
degenerate at the chiral crossover temperature indicating an effective
non-restoration of the axial symmetry. Also the splitting between the vector and the
pseudo-scalar screening masses persists even in the limit of zero lattice spacing and
at a moderately high temperature around $420$ MeV. In the charmonium sector our
investigation shows that the screening masses of the pseudo-scalar and the vector
charmonia are almost (within a few percent) equal to their zero temperature masses
for temperatures less than $300$ MeV. We also present results for the charmonium
screening masses using periodic boundary conditions along the temporal direction and
discuss their implications.

\end{abstract}

\end{frontmatter}
\section{Introduction}

In-medium properties of mesonic observables provide insight into the details of
structure and properties of the Quark Gluon Plasma (QGP). By studying these
properties one can obtain information concerning its important length-scales,
relevant degrees of freedom \etc. Such studies also help us to understand the nature
of the chiral and $U_A(1)$ axial symmetry restorations in Quantum ChromoDynamics
(QCD). Finite temperature lattice QCD simulation is, to date, the most viable and
successful technique for non-perturbative studies of such in-medium properties.
Since the maximum available physical temporal extent in a finite temperature lattice
QCD simulation is always limited by the inverse of the temperature it is easier to
use spatial correlation functions of meson-like excitations for the study of their
in-medium properties. Exponential decays of such spatial correlation functions define
the so-called \emph{screening masses} \cite{detar-1}. The inverse of the screening
mass indicates the typical distance beyond which the influence of a meson-like
excitation inside the QGP is effectively screened.

For the continuum non-interacting theory at a temperature $T$ the value of the
screening mass of a meson is given by \cite{florkowski-1} $M^s_{free}(T)=2\sqrt{(\pi
T)^2+m_q^2}$ , $m_q$ being the quark mass, independent of its spin-parity structure.
At very high temperatures and in the limit of zero quark mass perturbative
calculations \cite{pert-calc} show that this \emph{Free continuum} limit of
$M^s_{free}(T)/T=2\pi$ is reached from above. Although in general the screening mass
is not identical to the \emph{pole mass} $m_M$ both the spatial and the temporal
correlation functions depend on the same spectral density of the meson-like
excitation \cite{karsch-1}. This in turn ensures \cite{karsch-1} that for a stable
(free) mesonic state $M^s(T)=m_M$, \eg at $T=0$.

Mesonic screening masses have been computed in many different lattice QCD
simulations. For a review of earlier lattice results see \Ref{karsch-1}. Some recent
lattice results can be found in \Ref{latest-lat-res} and \Ref{laermann-1}. In this
work we present results for the mesonic screening masses from $2+1$ flavor lattice
QCD simulations with two degenerate dynamical up ($u$), down ($d$) quarks and a
dynamical strange ($s$) quark. In this work we have not treated the charm ($c$) quark
as a dynamical flavor. Since the mass of the $c$ quark is much larger than our
explored $T$ range such a partially-quenched treatment is expected to be a very good
approximation. For our simulations we have used an improved staggered fermion
(p4fat3) action with $3$ different extents in the temporal direction, \viz
$N_\tau=4,\ 6$ and $8$, keeping the extents of the spatial directions fixed at
$N_s=4N_\tau$. Quarks masses have been tuned such that $m_\pi\simeq220$ MeV,
$m_K\simeq500$ MeV, $m_\Psi\simeq3097$ MeV and $m_{\eta_c}\simeq2980$ MeV at zero
temperature. Further details of our simulations can be found in \Ref{rbc-bi} and
details on the tuning of the charm quark mass can be found in \Ref{cheng-1}. On these
gauge configurations we have analyzed screening masses using $8$ different local
staggered meson operators involving (isovector) \emph{scalar} ($SC$),
\emph{pseudo-scalar} ($PS$), (transverse) \emph{vector} ($V$) and (transverse)
\emph{axial-vector} ($AV$) channels for $4$ different combinations of quark fields,
\viz $\bar ud$, $\bar us$, $\bar ss$ and $\bar cc$.  More technical details on the
meson operators and our analysis can be found in \Ref{swagato-1}. 

\section{Results}

\bfig
\subfigure[]{ \label{fig:ps_uu} \includegraphics[scale=0.33]{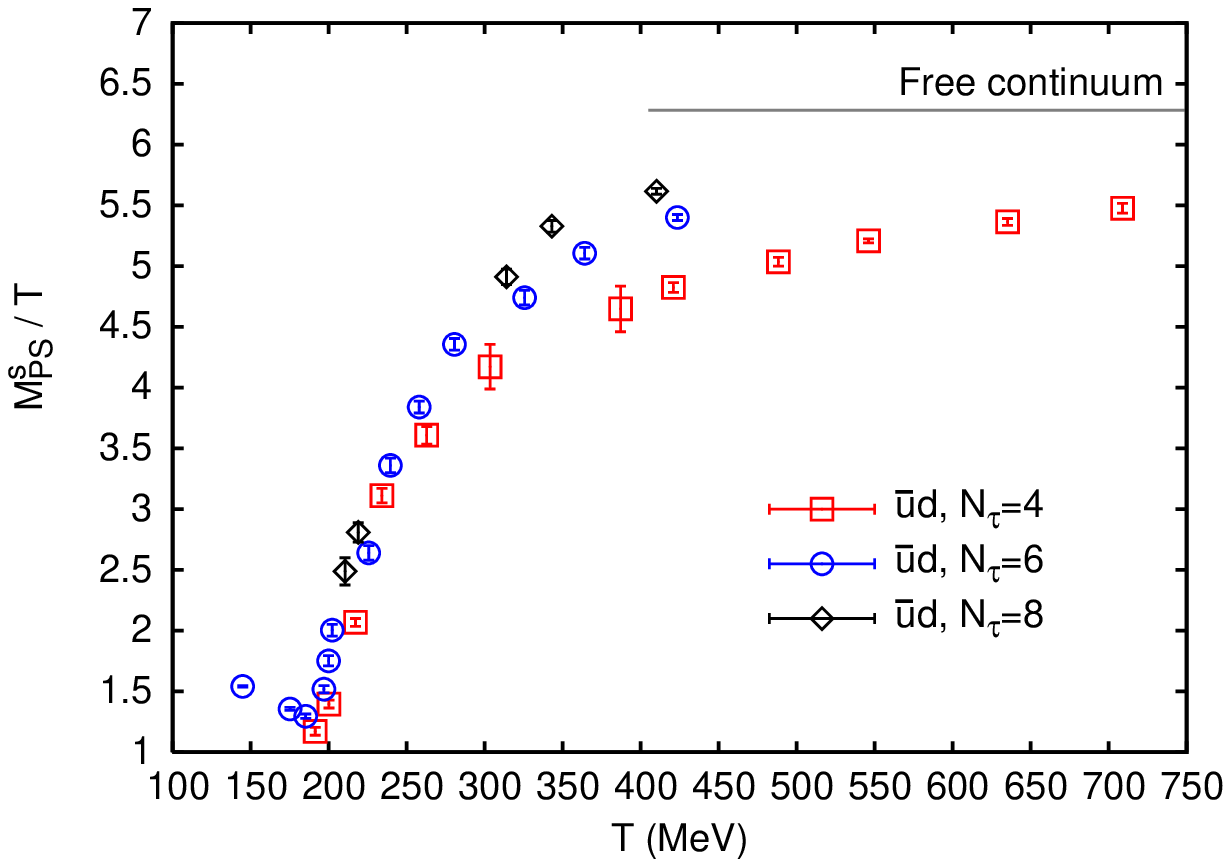} }
\subfigure[]{ \label{fig:sc_uu} \includegraphics[scale=0.33]{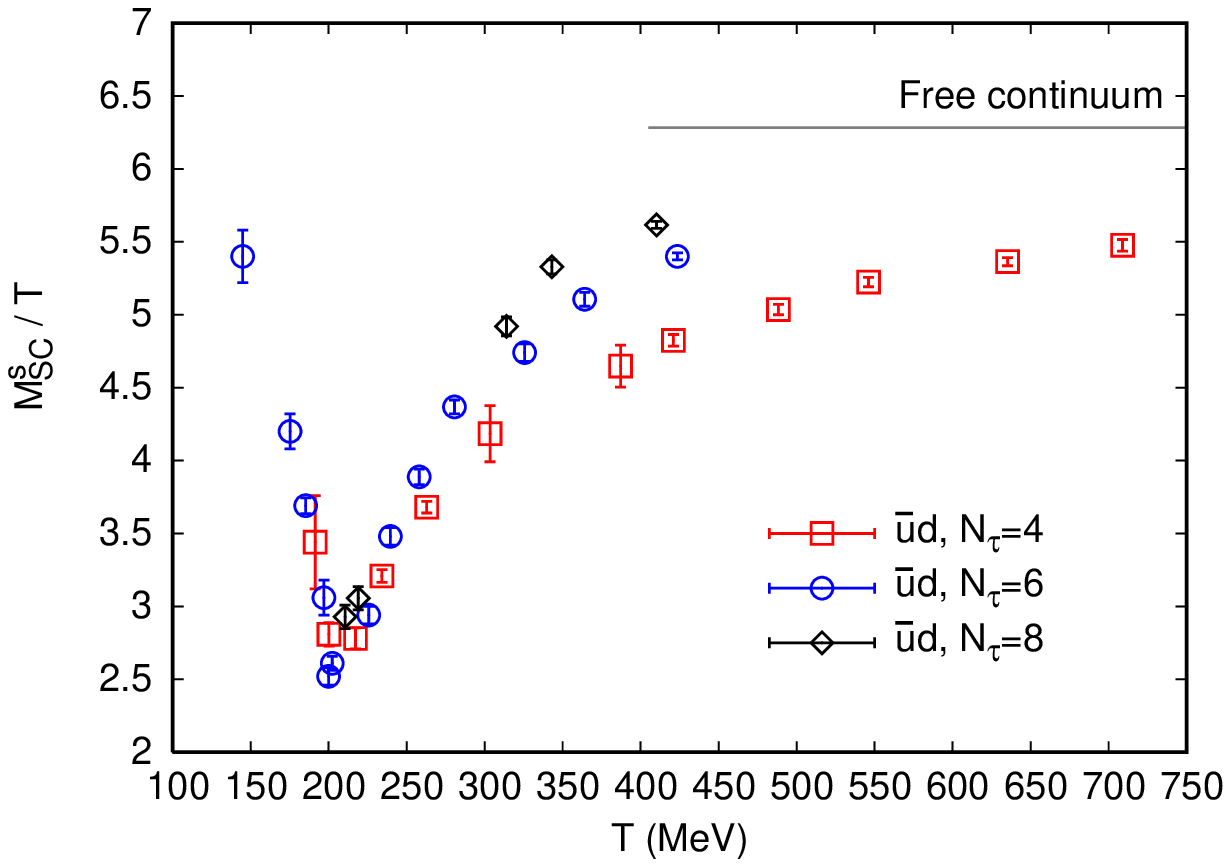} }
\subfigure[]{ \label{fig:v_uu} \includegraphics[scale=0.33]{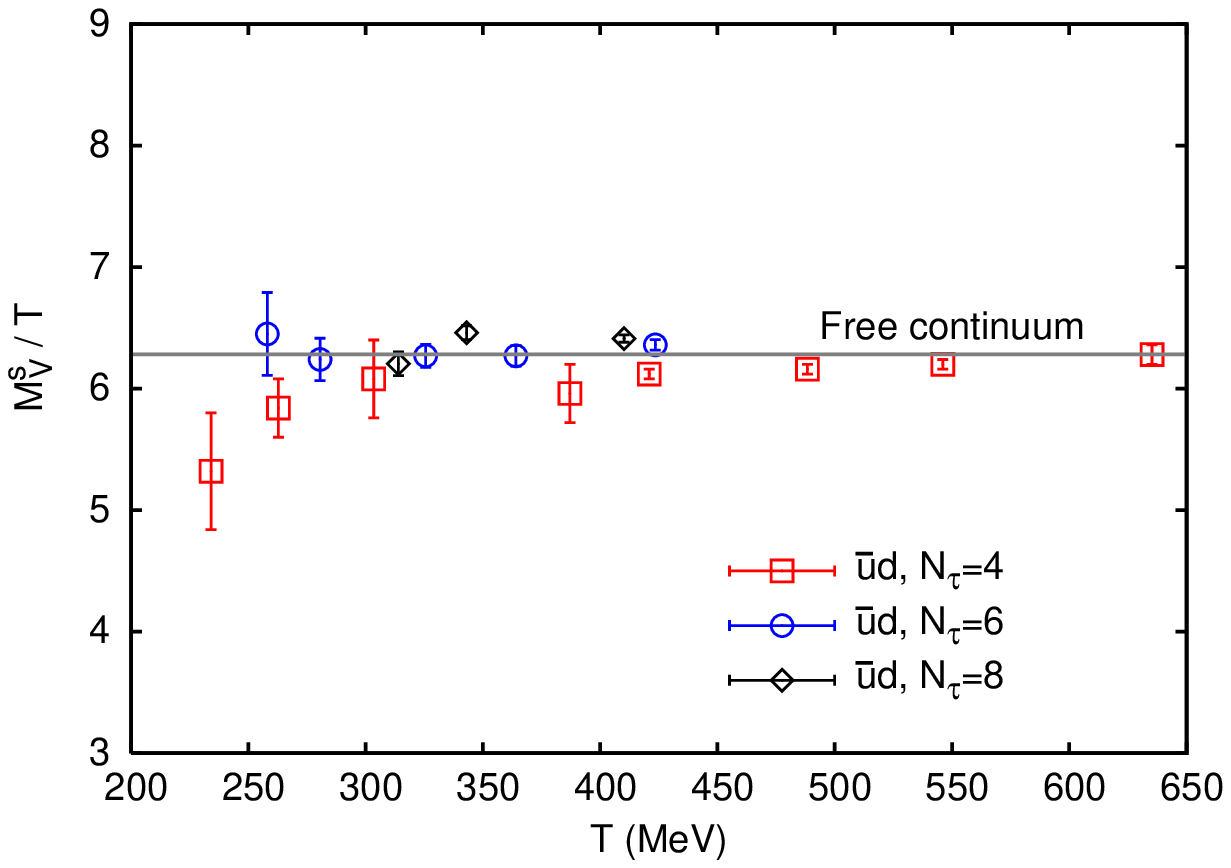} }
\subfigure[]{ \label{fig:av_uu} \includegraphics[scale=0.33]{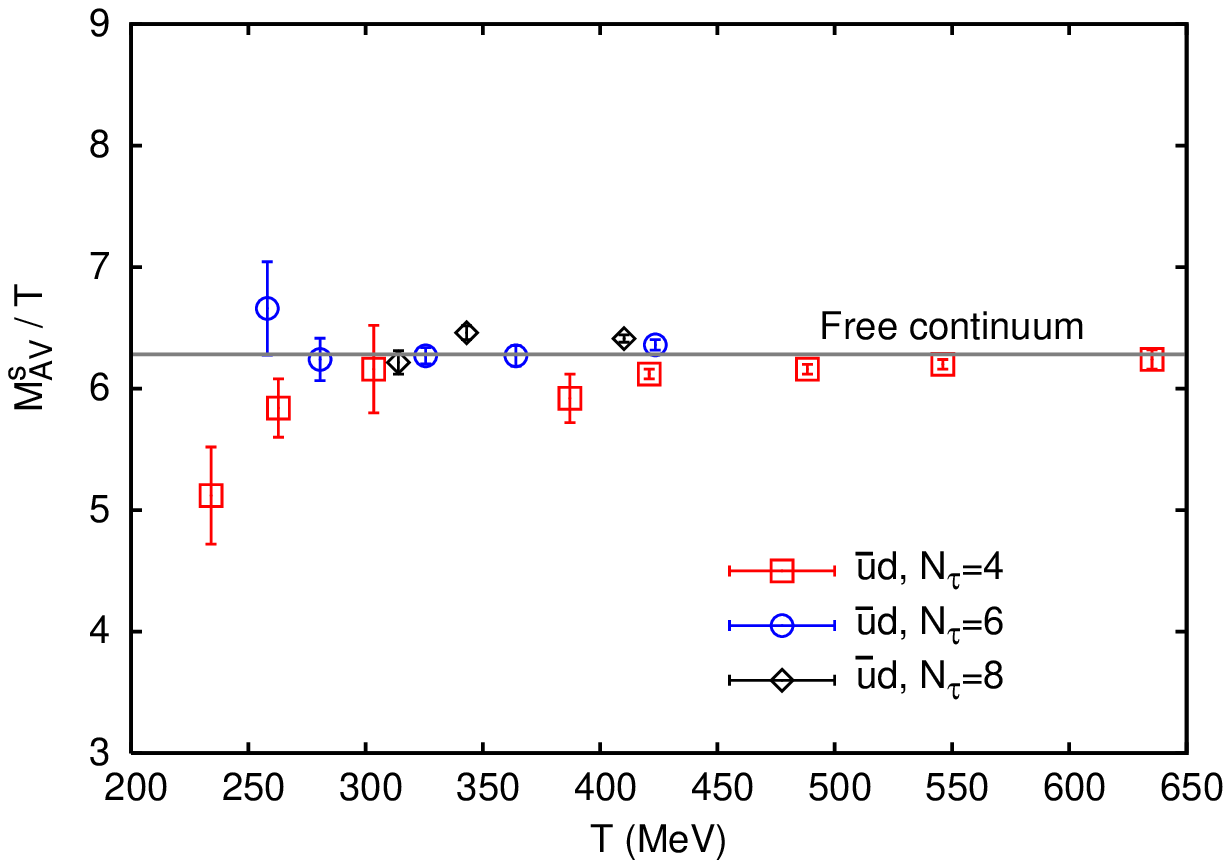} }
\subfigure[]{ \label{fig:sc-ps} \includegraphics[scale=0.33]{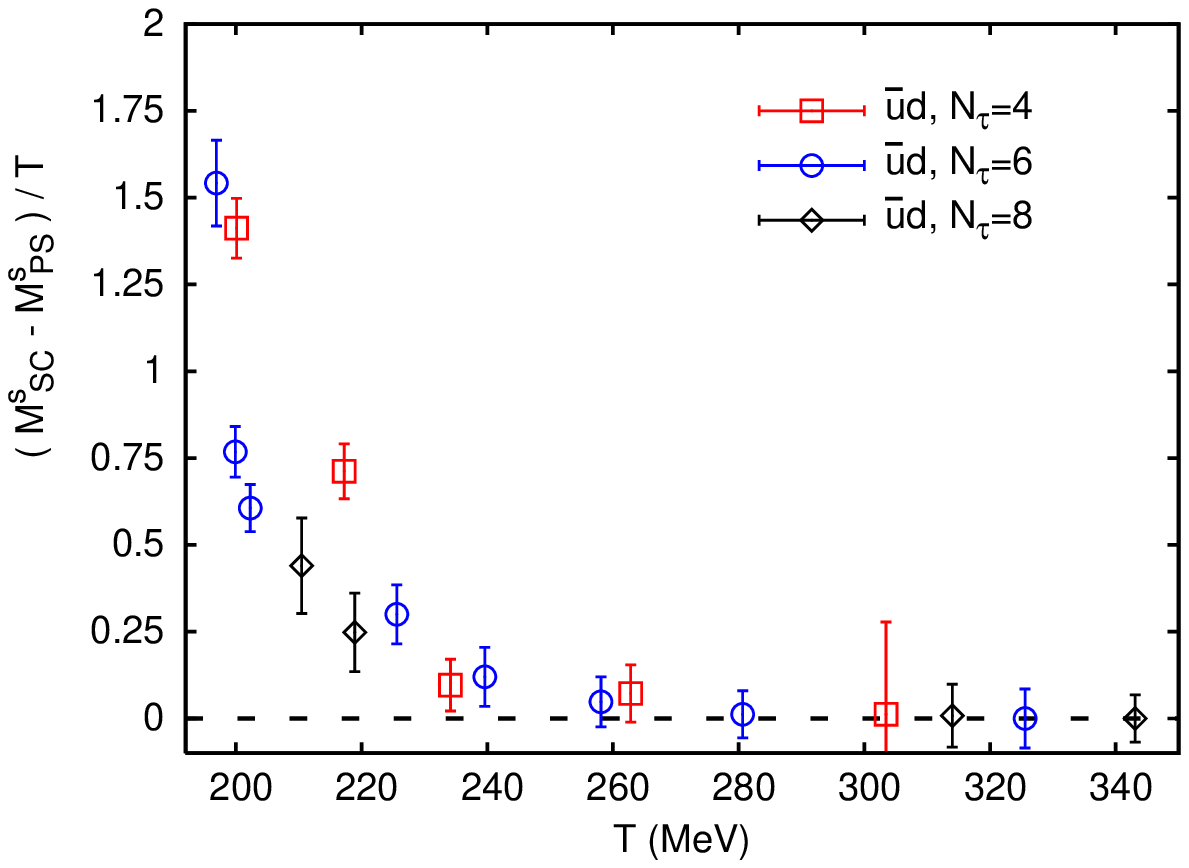} }
\subfigure[]{ \label{fig:extrap} \includegraphics[scale=0.33]{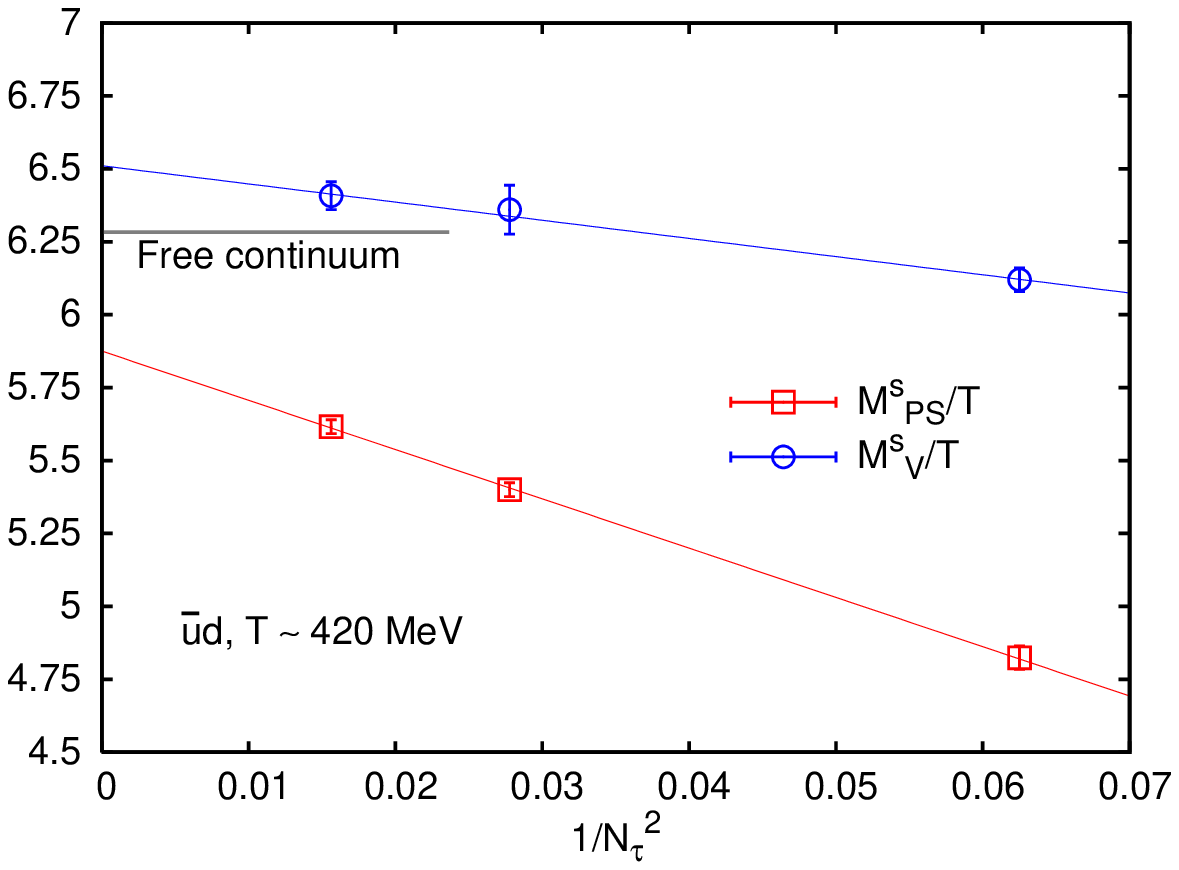} }

\caption{Screening masses in the $PS$ (a), $SC$ (b), $V$ (c) and $AV$ (d) channels
for the $\bar ud$ sector. (e) Difference between the screening masses in the $SC$ and
$PS$ channels. (f) Extrapolations of $M^s_V$ and $M^s_{PS}$ to zero lattice spacing
at $T\simeq420$ MeV.}

\label{fig:scr_mass_uu}
\efig

In \fig{fig:scr_mass_uu} we show our results for the screening masses in the $\bar
ud$ sector for the $4$ different mesonic channels. Upto the chiral crossover
temperature, $T_{pc}\simeq 190$ MeV, $M^s_{PS}$ remains approximately equal to
$m_\pi$ but $M^s_{SC}$ shows a distinct minimum around $T_{pc}$. For $T>T_{pc}$
screening masses in both these channels grow rapidly but stay well below the
\emph{Free continuum} limit of $2\pi T$ even at $T\simeq420$ MeV. On the other hand,
for $T\gtrsim T_{pc}$ within our errors we found that $M^s_V\simeq M^s_{AV} \simeq
2\pi T$. The behavior of screening masses in the all four channels for the $\bar us$
and $\bar ss$ sectors are qualitatively similar to that of the $\bar ud$ sector. 

Above the chiral crossover temperature chiral symmetry gets restored and hence the
$V$ and $AV$ are expected to become degenerate. This is exactly what we have found in
our analysis of the screening masses in those two channels. On the contrary, as can
be seen from \fig{fig:sc-ps}, the screening masses in the pseudo-scalar and the
(isovector) scalar channels do not become degenerate in the temperature range
$T_{pc}<T\lesssim250$ MeV. This observation indicates that the effective restoration
of the $U_A(1)$ axial symmetry does not coincide with the chiral symmetry restoration
and takes place at a temperature $T>T_{pc}$.

We have also investigated the cut-off dependence in the meson screening masses. In
\fig{fig:extrap} we show the continuum limit extrapolation (\ie a linear
extrapolation in the lattice spacing squared $a^2=1/(TN_\tau)^2\to0$) of $M^s_V$ at a
temperature $T\simeq420$ MeV. Such an extrapolation shows that even at this
`not-so-high' temperature $M^s_V>2\pi T$ in accordance with the perturbative
predictions \cite{pert-calc}. However, as observed in \Ref{laermann-1} for
simulations with quenched Wilson fermions, this overshooting of the free continuum
value can very well be a finite-volume effect and may go away in the infinite volume
limit. The cut-off effects are more pronounced in the $PS$ and $SC$ channels
(\fig{fig:ps_uu} and \fig{fig:sc_uu}). A similar continuum limit extrapolation, at
$T\simeq420$ MeV, of $M^s_{PS}$ shows that the extrapolated value of $M^s_{PS}$
remains well below the free continuum limit and the splitting between $M^s_V$ and
$M^s_{PS}$ persists even in the limit of zero lattice spacing (see \fig{fig:extrap}).

As discussed before, for a stable (free) mesonic state the screening mass and the
pole mass are identical. With this hindsight one may compare the screening mass of a
charmonium to its zero temperature pole mass to shed some light on the issue of
survival/dissociation a charmonium in QGP. We have analyzed the mesonic screening
masses in $PS$ and $V$ channels for the heavy-quark $\bar cc$ sector and compared
them with the $T=0$ pole masses in the corresponding channels. Amazingly, the ratios
$M^s_{PS}/m_{\eta_c}$ (\fig{fig:ps_cc}) and $M^s_V/m_\Psi$ (\fig{fig:v_cc}) stay very
close to $1$ (within $<5\%$) till $T\lesssim300$ MeV. Beyond this temperature these
ratios seem to increase rapidly as the temperature increases. In order to illustrate
the distinctiveness of the $\bar cc$ sector more clearly in \fig{fig:ratio} we have
plotted the ratios of the $PS$ screening masses to the corresponding zero temperature
pole masses for all the four different combinations of quark fields on $6\times24^3$
lattices. While around $T\simeq275$ MeV there is almost no difference in the
screening and the pole mass in the $\bar cc$ sector the relative difference is
$\sim90\%$ even for the $\bar ss$ sector.

\bfig
\subfigure[]{ \label{fig:ps_cc} \includegraphics[scale=0.33]{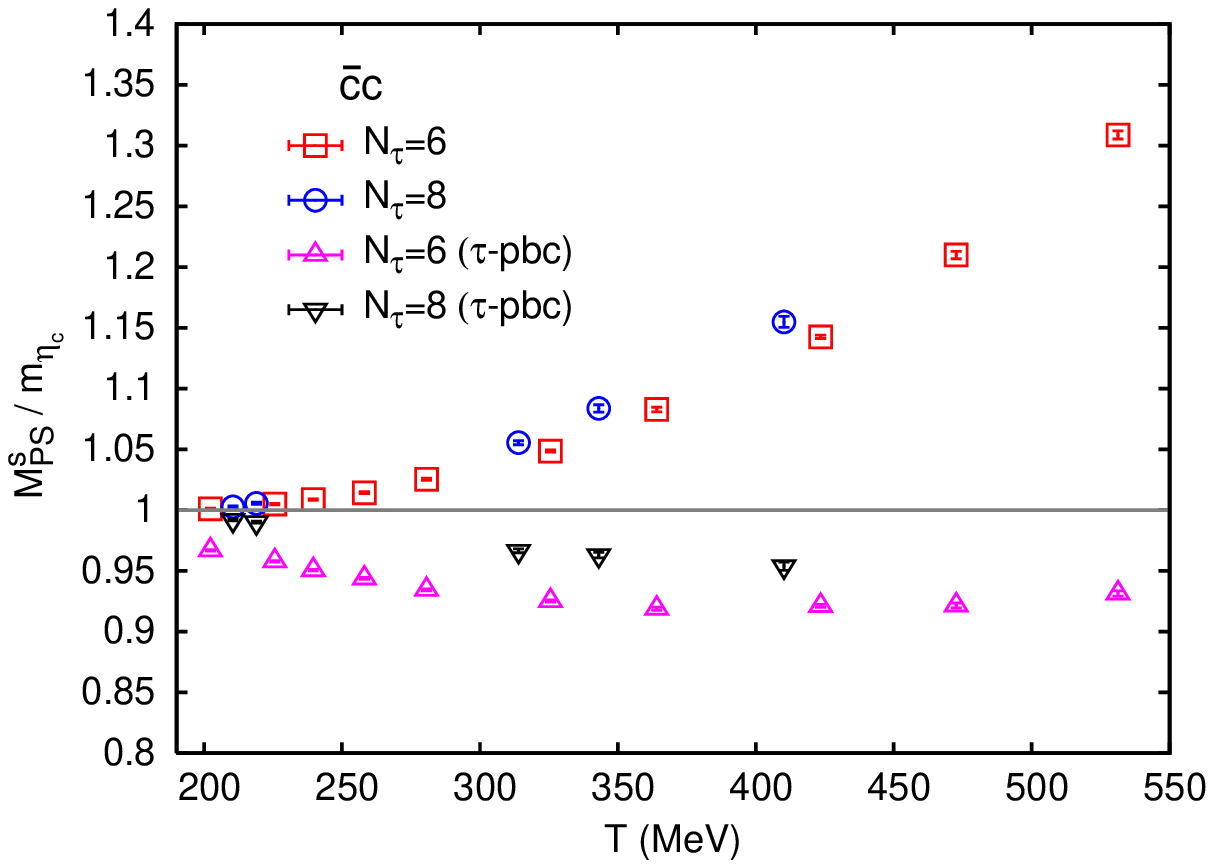} }
\subfigure[]{ \label{fig:v_cc} \includegraphics[scale=0.33]{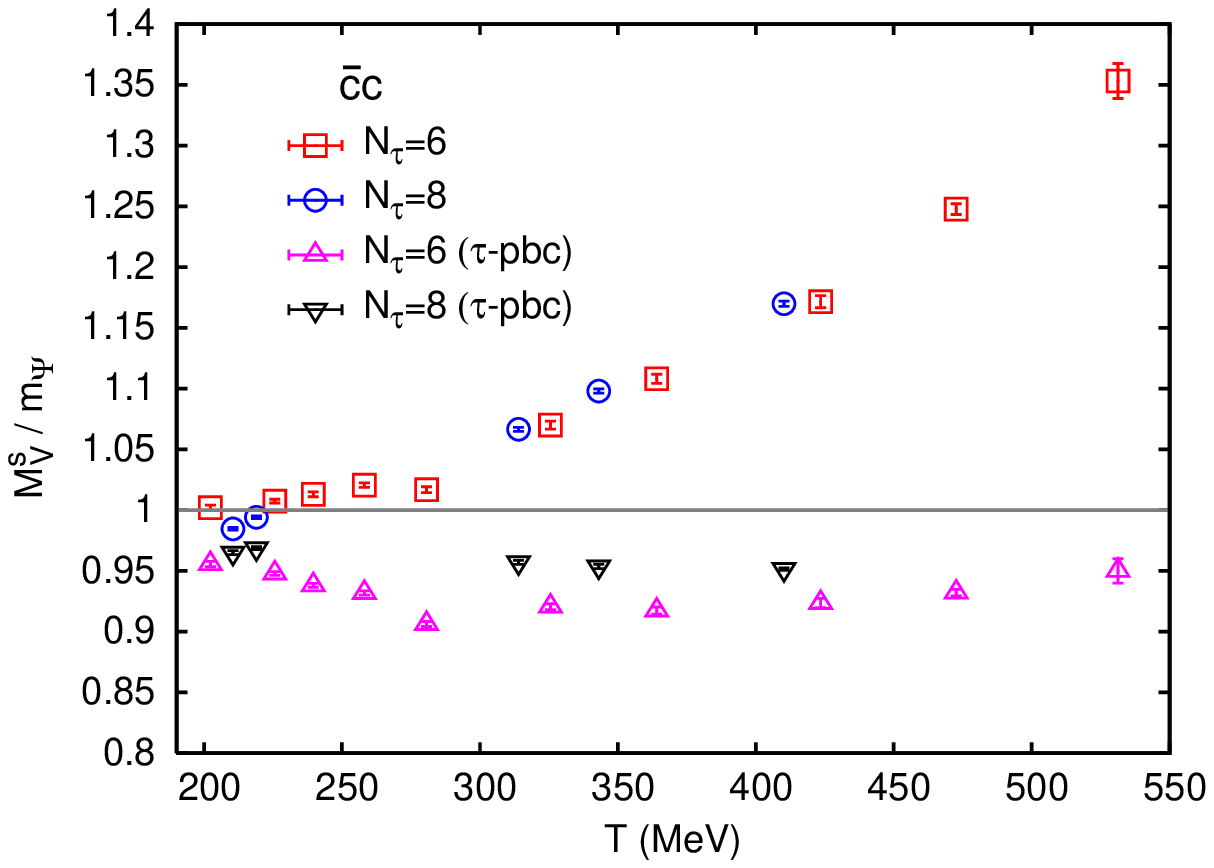} }
\subfigure[]{ \label{fig:ratio} \includegraphics[scale=0.33]{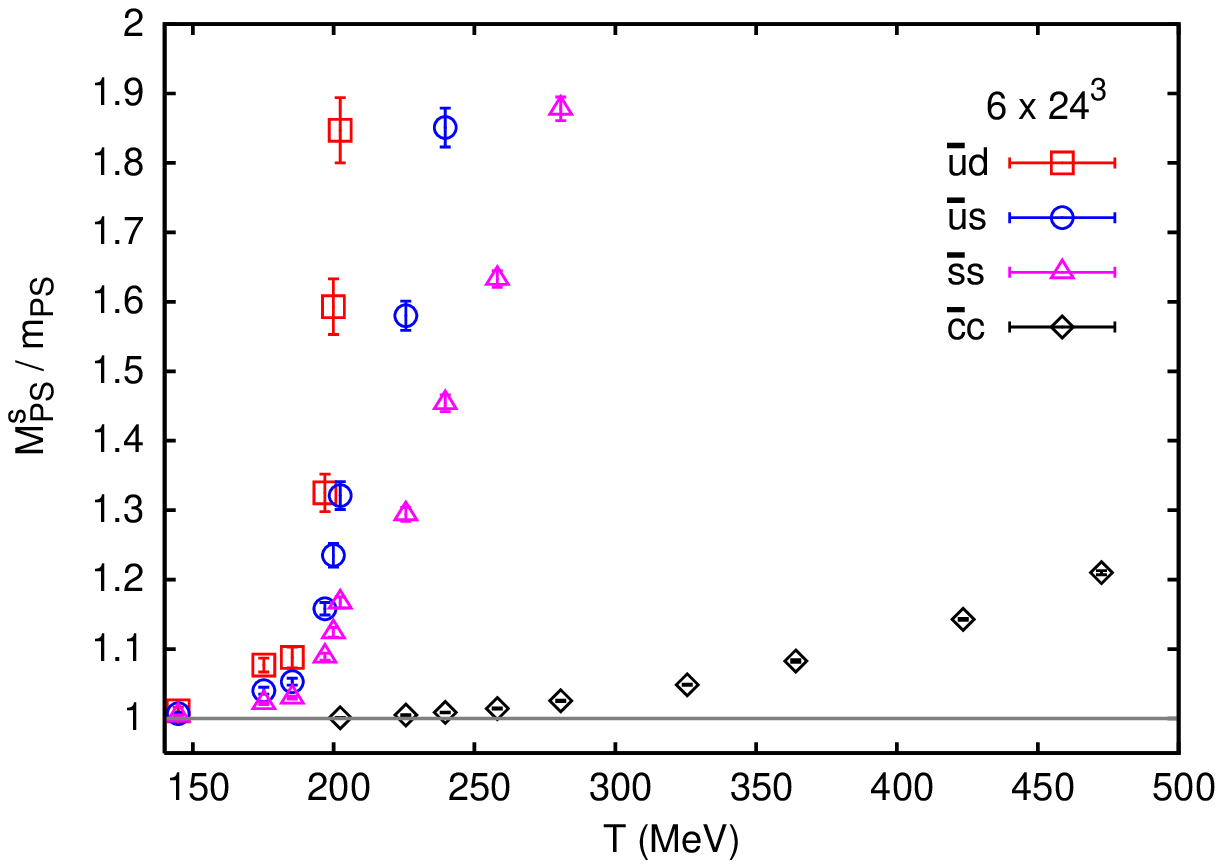} }

\caption{Ratio of the screening mass to the zero temperature pole mass in $PS$ (a)
and $V$ (b) channels for the $\bar cc$ sector. \emph{$\tau$-pbc} means screening
masses have been extracted by employing periodic boundary conditions along the
temporal direction. (c) The same ratio in the $PS$ channel for four different
anti-quark---quark sectors using $6\times24^3$ lattices.}
\efig

For a deeper understanding of these results we have also analyzed the screening
masses in these channels by employing periodic, in contrast to the standard
anti-periodic, boundary conditions (denoted by \emph{$\tau$-pbc}) along the temporal
direction. If one employs \emph{$\tau$-pbc} then in the free (continuum) case the
contribution to the mesonic screening mass will come from the lowest bosonic
Matsubara frequency ($\omega_{min}=0$), as oppose to the lowest fermionic Matsubara
frequency ($\omega_{min}=\pi T$), and consequently $M^s_{free,\tau-pbc}(T)=2m_q$
independent of $T$. Thus a large splitting between $M^s(T)$ and $M^s_{\tau-pbc}(T)$
will indicate that the contribution to screening mass at that temperature is coming
from two quasi-quarks.  On the other hand, for a free stable mesonic state one may
expect that $M^s(T)\simeq M^s_{\tau-pbc}$. In \fig{fig:ps_cc} and \fig{fig:v_cc} we
have compared our results of $M^s_{\tau-pbc}(T)$ with the corresponding $M^s(T)$.
While the difference between $M^s$ and $M^s_{\tau-pbc}$ is $\gtrsim20\%$ for
$T\gtrsim300$ MeV the difference is only at the level of a few percent around
$T\simeq220$ MeV. These relatively small differences at lower temperatures are even
reduced as one decreases the lattice spacing from $aT=1/6$ to $1/8$, although the
qualitative picture for $T\gtrsim300$ MeV remains almost unchanged. All these
findings are very interesting and seem to suggest that $\eta_c$ and $J/\Psi$ may
survive in a QGP at least for temperatures less than $1.5T_{pc}$.




\begin{thebibliography}{00}

\bibitem{detar-1}
C.\ DeTar and J.\ B.\ Kogut, \prl{59}{399}{1987}; \ibid \prd{36}{2828}{1987}.

\bibitem{florkowski-1}
W.\ Florkowski and B.\ L.\ Friman, Z. Phys. {\bf A347}, (1994) 271.

\bibitem{pert-calc}
M.\ Laine and M.\ Vepsalainen, \jhep{02}{004}{2004}; W.\ M.\ Alberico \etal, Nucl.\
Phys.\ {\bf A792}, (2007) 152.

\bibitem{karsch-1}
F.\ Karsch and E.\ Laermann, in Quark Gluon Plasma {\bf III} (R.\ Hwa ed.)
[hep-lat/0305025].

\bibitem{latest-lat-res}
R.\ V.\ Gavai, S.\ Gupta and R.\ Lacaze, arXiv:0803.1368 [hep-lat], I.\ Pushkina
\etal\ (QCD-TARO collaboration), \plb{609}{265}{2005}.

\bibitem{laermann-1}
E.\ Laermann \etal, PoS {\bf LAT2008}, (2008) 193.

\bibitem{rbc-bi}
M.\ Cheng \etal, \prd{77}{014511}{2008}.

\bibitem{cheng-1}
M.\ Cheng, PoS {\bf LAT2007}, (2007) 173. 

\bibitem{swagato-1}
S.\ Mukherjee, PoS {\bf LAT2007}, (2007) 210.
 

\end{thebibliography}
\end{document}